\documentclass[reprint]{revtex4}
\usepackage{amsmath}
\usepackage{graphicx}
\usepackage{dcolumn}
\usepackage{subfigure}
\usepackage{bm}

\begin{document}

\title{Quatnum coherence behaviors of fermionic system in non-inertial frame}

\author{Zhiming Huang}
\email{465609785@qq.com}
\affiliation{School of Economics and Management, Wuyi University, Jiangmen 529020, China}
\author{Haozhen Situ}
\email{situhaozhen@gmail.com}
\affiliation{College of Mathematics and Informatics, South China Agricultural University, Guangzhou 510642, China}
\author{Cai Zhang}
\email{zhangcai.sysu@gmail.com}
\affiliation{College of Mathematics and Informatics, South China Agricultural University, Guangzhou 510642, China}

\begin{abstract}
In this paper, we analyse the quantum coherence behaviors of a single qubit in the relativistic regime beyond the single-mode approximation.
Firstly, we investigate the freezing condition of quantum coherence in fermionic system. We also study the quantum coherence tradeoff between particle and antiparticle sector. It is found that there exists quantum coherence transfer between particle and antiparticle sector, but the coherence lost in particle sector is not entirely compensated by the coherence generation of antiparticle sector. Besides, we emphatically discuss the cohering power and decohering power of Unruh channel with respect to the computational basis. It is shown that cohering power is vanishing and decohering power is dependent of the choice of Unruh mode and acceleration. Finally, We compare the behaviors of quantum coherence with geometric quantum discord and entanglement in relativistic setup. Our results show that this quantifiers in two region converge at infinite acceleration limit, which implies that this measures become independent of Unruh modes beyond the single-mode approximations. It is also demonstrated that the robustness of quantum coherence and geometric quantum discord are better than entanglement under the influence of acceleration, since entanglement undergoes sudden death.
\end{abstract}
\maketitle

\section{Introduction}
Relativistic quantum information \cite{Peres2004,Friis2012,Bruschi2012,Wang2016,Wang2016a} has attracted much attention in recent years, which is the integration of relativity theory, quantum field theory, and quantum information theory.
%Unruh effect is an important prediction of quantum field theory, which tells us that the noninertial observers that move with a uniform acceleration in Minkowski spacetime see the Minkowski vacuum of quantum fields as a thermal bath \cite{Unruh1976,Crispino2008}. The Unruh effect indicates that the concept of particle is observer-dependent and relative.
%$Studying quantum correlations in a relativistic framework is necessary because the world is essentially noninertial \cite{Wang22010}.
It is believed that the study of quantum correlation in a relativistic setting is not only helpful in understanding some key questions in quantum information theory, but also plays an important role in the study of quantum effects of black holes \cite{Wang22010,Martin2009,Martin2010,Martin22010}.
Many authors \cite{Alsing2006,Pan2008,Hwang2010,Wang2016b,Montero2011,Ramzan2012,Wang32010,Brown2012,Ramzan2013} have studied the behaviors of entanglement and discord-type quantum correlations in fermionic system of non-inertial frames based on single-mode approximation (SMA). %Some papers \cite{} investigated the discord-type quantum correlations in fermionic system using SMA. Most analysis for the fermionic systems was only based on the SMA, however, to obtain a precise understanding about quantum correlation dynamic, one has to give a full consideration beyond SMA\cite{Bruschi2010}.
 Recently, Some papers \cite{Bruschi2010,Montero22011,Martin2011,Montero2012,Chang2012,Ramzan2014} consider the behaviors of quantum correlations of fermionic system in accelerated frame beyond SMA. On the other hand, Quantum coherence resulting from quantum state superpositon is a fundamentally important physical resource, which plays a key role in quantum physics and quantum information processing, such as quantum optics \cite{Glauber1963,Sudarshan1963,Scully1991,Mandel1995,Asboth2005,Vogel2014,Mraz2014}, thermodynamics \cite{RoBnagel2014,Aberg2014,Correa2014,Narasimhachar2015,Lostaglio2015}, quantum information \cite{Nielsen2000,Baumgratz2014}, solid state physics \cite{Deveaud2009,Li2012}, and quantum biology \cite{Engel2007,Collini2010,Lambert2013,Chin2013,Cai2013}.

Despite the prominent role of quantum coherence, only very recently has a theoretic framework to measure coherence for quantum states been developed \cite{Baumgratz2014,Levi2014,Girolami2014,Smyth2014,Pires2015,Streltsov2015,Rana2016}. Ref. \cite{Baumgratz2014} proposed the rigorous characterization of coherence in the framework of resource theory which is based on identifying the set of incoherent states and a class of incoherent operations. Based on such a framework, one can define suitable measures, including the relative entropy of coherence and $l_1$ norm of coherence \cite{Baumgratz2014}, and the skew information \cite{Girolami2014}. According to resource theory, entanglement and other types of quantum correlations beyond entanglement, for instance, quantum discord, were discovered as key quantum resources. Resources can often be traded for another. Recently, some papers \cite{Streltsov2015,Yao2015,Chitambar2015} explored the relation between coherence and entanglement. Some authors \cite{Yao2015,Ma2016,Xi22015,Hu2015} studied the link between coherence and discord-type quantum correlation. Very recently, some work \cite{Mani2015,Garc¨ªa2015,Xi2015} introduced and investigated the cohering and decohering power of quantum channels. The cohering power and decohering power of a quantum channel are the power of a quantum channel creating or destroying the coherence of input quantum states, which is a fundamental concept within the framework of the resource theory of coherence.

%As far as we know, that many papers discuss quantum correlations behavior is based on bipartite fermionic system \cite{Bruschi2010,Montero22011,Martin2011,Montero2012,Chang2012,Ramzan2014}.
In this paper, we explore the quantum coherence behavior for one party fermionic system in accelerated frame beyond SMA. As mentioned above, quantum coherence as a crucial quantum resource is closely connected with other nonclassical correlation resource, thus we will compare the dynamics of quantum coherence in an accelerated frame with entanglement and geometric quantum discord. We find that quantum coherence, geometric quantum discord and entanglement have different behavior with other nonclassical correlation \cite{Bruschi2010,Martin2011,Chang2012,Ramzan2014}. Actually, the acceleration of observers can be treated as certain \lq\lq environmental decoherence\rq\rq\ and the effect will degrade the quantum correlation from the perspective of accelerated observers \cite{Alsing2006}. Therefore, Unruh effect can be regarded as a kind of decoherence channel, and we will study cohering and decohering power in non-inertial frame beyond SMA.  We believe that such explorations will deepen our understanding of quantum coherence in relativistic systems and is of significance from the point of view of both fundamental theory and practical applications.

The organization of this paper is as follows. Sec.\ref{S2} introduces fermionic system in noninertial frame beyond SMA. In Sec.\ref{S3}, we recall some concepts about quantum coherence and quantum correlation. Sec.\ref{S4} studies the behaviors of quantum coherence, cohering and decohering power in an accelerated frame. A brief conclusion is given in the last sections.
\section{Noninertial frame}\label{S2}
An observer with constant proper acceleration is best described Rindler coordinate $(\tau,\zeta,y,z)$, instead of the Minkowski coordinate $(t, x, y, z)$. Rindler coordinate and Minkowski coordinate have the following relation:
\begin{align}
ct=\zeta\sinh(\frac{a\tau}{c}),\qquad x=\zeta\cosh(\frac{a\tau}{c}),\label{eq1}
\end{align}
\begin{figure}
\centering
\includegraphics[height=4cm,width=4cm]{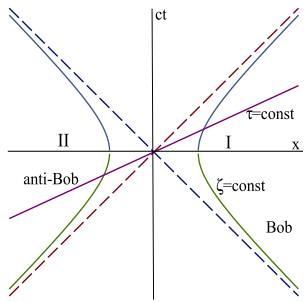}
\caption{\label{figure1}Rindler space-time diagram: Bob is in the uniformly accelerated frame. Two regions I and II are causally disconnected.}
\end{figure}
where the curve $\zeta = 1/a$ is the world line of a uniformly accelerated observer, and $a$ is equal to the proper acceleration of uniformly accelerated observer \cite{Martin2011}. $c$ is the velocity of light. For fixed $\zeta$, the coordinate can describe hyperbolic trajectories in spacetime (see Fig. \ref{figure1}). In fact, Eq. (\ref{eq1}) only covers region I. Region II is covered by $ct=-\zeta\sinh(\frac{a\tau}{c}), x=-\zeta\cosh(\frac{a\tau}{c})$. The other two regions can be covered with $ct=\pm\zeta\cosh(\frac{a\tau}{c}), x=\pm\zeta\sinh(\frac{a\tau}{c})$.

The Dirac field $\phi$ satisfies equation $\{i\gamma^{\mu}((\partial_{\mu}-\Gamma_{\mu})+m\}=0$, where $\gamma^{\mu}$ are the Dirac-Pauli matrices and $\Gamma_{\mu}$ are the spin connection coefficients. In terms of Minkowski modes and Rindler modes, the Dirac field is given by
\begin{align}
\phi&=N_{M}\sum_{i}(a_{i,M}\upsilon_{i,M}^{+}+b_{i,M}^{\dag}\upsilon_{i,M}^{-})\nonumber\\
&=N_{R}\sum_{j}(a_{j,I}\upsilon_{j,I}^{+}+b_{j,I}^{\dag}\upsilon_{j,I}^{-}+a_{j,II}\upsilon_{j,II}^{+}+b_{j,II}^{\dag}\upsilon_{j,II}^{-}),
\end{align}
where $N_{M}$ and $N_{R}$ are normalization constants. $\upsilon_{i,M}^{\pm}$ denotes positive and negative energy solutions of the Dirac equation in Minkowski coordinate, which can be gotten with respect to the Killing vector field in Minkowski spacetime. $\upsilon_{i,I}^{\pm}$ and $\upsilon_{i,II}^{\pm}$ are the positive and negative energy solutions of the Dirac equation in Rindler coordinate, with respect to the timelike Killing vector field in regions I and II. $a_{i,\sigma}^{\dag} (a_{i,\sigma})$ and $b_{i,\sigma}^{\dag} (b_{i,\sigma})$ are the the creation (annihilation) operators for the positive and negative energy solutions (particle and antiparticle), where $\sigma=\{M,I,II\}$, and satisfy the usual anticommutation rule
\begin{align}
\{a_{i,\sigma},a_{j,\sigma^{\prime}}^{\dag}\}=\{b_{i,\sigma},b_{j,\sigma^{\prime}}^{\dag}\}=\delta_{ij}\delta_{\sigma\sigma^{\prime}}.
\end{align}
Linear combinations of Rindler modes constructing Unruh modes, each Unruh mode can be transformed to a monochromatic Rindler mode. This transformation is given by \cite{Bruschi2010}:
\begin{align}
A_{i,R/L}\equiv \cos r_{i}a_{i,I/II}-\sin r_{i}b_{i,II/I}^{\dag}
\end{align}
where $\cos r_{i}=(e^{\frac{-2\pi\Omega c}{a}}+1)^{-1/2}$, and $\Omega$ denotes the Rindler frequency. However, a more general relation can be obtained,
\begin{align}
a_{i,U}^{\dag}=&q_{L}(A_{i,L}^{\dag}\otimes I_{R})+q_{R}( I_{L}\otimes A_{i,R}^{\dag}),\nonumber\\
&q_{L}^{2}+q_{R}^{2}=1.
\end{align}
We assume that $q_{L}$ and $q_{R}$ are real numbers. By the above equation, one can obtain Unruh mode beyond the single-mode approximation. The case $q_{R}=1$ corresponds to the single-mode approximation.

Grassman scalars field, which is an anticommuting field with only one degree of freedom and is the the simplest case of Dirac field, preserves the fundamental Dirac characteristics. Using the above relations, Unruh vacuum for the Grassman scalar field is given by
\begin{align}
|0_{\Omega}\rangle_{U}=&\cos^{2}r_{\Omega}|0000\rangle_{\Omega}-\sin r_{\Omega}\cos r_{\Omega}|0011\rangle_{\Omega}\nonumber\\
&+\sin r_{\Omega} \cos r_{\Omega}|1100\rangle_{\Omega}-\sin^{2} r_{\Omega}|1111\rangle_{\Omega},\label{U0}
\end{align}
where the modes labeled with $U$ are Grassman-Unruh modes.
Here, we introduce the notation $|pqmn\rangle_{\Omega}\equiv|p_{\Omega}\rangle^{+}_{I}|q_{\Omega}\rangle^{-}_{II}|m_{\Omega}\rangle^{-}_{I}|n_{\Omega}\rangle^{+}_{II}$. The one-particle state is given by
\begin{align}
|1_{\Omega}\rangle_{U}^{+}=&q_{R}(\cos r_{\Omega}|1000\rangle_{\Omega}-\sin r_{\Omega}|1011\rangle_{\Omega})\nonumber\\
&+q_{L}(\sin r_{\Omega} |1101\rangle_{\Omega}+\cos r_{\Omega}|0001\rangle_{\Omega}), \label{U1}
\end{align}
\begin{align}
|1_{\Omega}\rangle_{U}^{-}=&q_{L}(\cos r_{\Omega}|0100\rangle_{\Omega}-\sin r_{\Omega}|0111\rangle_{\Omega})\nonumber\\
&+q_{R}(\sin r_{\Omega} |1110\rangle_{\Omega}+\cos r_{\Omega}|0010\rangle_{\Omega}).\label{U2}
\end{align}
The states labeled by $\Omega$ are Unruh states, and we will omit the label $\Omega$ for simplicity in the rest of the present paper. Here, we will apply the physical structure suggested by Ref. \cite{Montero2012,Chang2012} that the ordering of the fermionic system should be rearranged by the sequence of particles and antiparticles in the separated region.

\section{Basic definitions}\label{S3}
Baumgratz et al. \cite{Baumgratz2014} introduced a rigorous framework for the quantification of coherence and identified easily computable measures of coherence.
By fixing a particular basis $\{|i\rangle\}$ in the $d$-dimensional Hilbert space, all density
operators of the form
\begin{align}
\hat{\delta}=\sum_{i=1}^{d}p_{i}|i\rangle\langle i|
\end{align}
are called incoherent states, and we label this set of quantum states by $\mathcal{I}$.
A quantum operation $\rho\rightarrow\sum_n K_n\rho K_n^\dagger$ is called an incoherent operation if the condition $K_n \mathcal{I} K_n^\dagger \subset \mathcal{I}$ is satisfied for all $n$.
A reasonable measure of quantum coherence $C$ should satisfy the following conditions in line with the resource theory \cite{Baumgratz2014}:
\begin{enumerate}
\item $C(\rho)=0$ iff $\rho\in \mathcal{I}$.
\item Monotonicity under non-selective incoherent completely positive and trace preserving (ICPTP) maps: $C(\rho)\geq C(\Phi_{\text{ICPTP}}(\rho))$, where $\Phi_{\text{ICPTP}}(\rho)=\sum_n K_n\rho K_n^\dagger$ and $\{K_{n}\}$ is a set of Kraus operators with $\sum_n K_n^\dagger K_n=I$ and $K_n \mathcal{I} K_n^\dagger \subset \mathcal{I}$.
\item Monotonicity under selective measurements on average: $C(\rho)\geq\sum_n p_n C(\rho_n)$, where $\rho_{n}= K_n\rho K_n^\dagger/p_{n}$, $p_{n}=\text{Tr}(K_n\rho K_n^\dagger)$ for all $\{K_{n}\}$ with $\sum_n K_n^\dagger K_n=I$ and $K_n \mathcal{I} K_n^\dagger \subset \mathcal{I}$
\item Convexity: $\sum_n p_n C(\rho_n)\geq C(\sum_n p_n \rho_n)$ for any set of states $\{\rho_n\}$ with probability $p_{n}\geq 0$ and $\sum_n p_n=1$.

\end{enumerate}
There are several coherence measures that satisfy the above conditions, such as the relative entropy of coherence and $l_1$ norm of coherence \cite{Baumgratz2014}.

Coherence properties of a quantum state are usually attributed to the off-diagonal elements of its density matrix with respect to a selected reference basis.
$l_1$ norm of coherence is a very intuitive quantification of coherence related to the off-diagonal elements of the considered quantum state, given by
\begin{align}
C_{l_{1}}(\rho)=\sum_{\substack{i\ne j}}|\rho_{i,j}|.\label{l1}
\end{align}

 The cohering power of a channel is the maximum amount of coherence that it creates when acting on  completely incoherent states. For a quantum channel $\mathcal{E}$, the cohering power is defined as \cite{Mani2015}
\begin{align}
\mathcal{C}_{\mathbf{k}}(\mathcal{E})=\max_{\rho\in\mathcal{I}}\{C_{\mathbf{k}}(\mathcal{E}(\rho))-C_{\mathbf{k}}(\rho)\}=\max_{\rho\in\mathcal{I}}C_{\mathbf{k}}(\mathcal{E}(\rho)),
\end{align}
where $C$ is any coherence measure and $\mathbf{k}$ denotes the observable basis. For a qubit channel $\mathcal{E}$, cohering power can be calculated by \cite{Mani2015}
\begin{align}
\mathcal{C}_{\mathbf{k}}(\mathcal{E})=\max\{F_{\mathcal{E}}(\mathbf{k},\mathbf{k}),F_{\mathcal{E}}(\mathbf{-k},\mathbf{k})\},\label{CP}
\end{align}
where
\begin{align}
F_{\mathcal{E}}(\mathbf{m},\mathbf{k})=(1-\sqrt{1-m^{\prime 2}})[1-(\frac{\mathbf{m^{\prime}}}{m^{\prime}}\cdot\mathbf{k})^{2}],\label{FC}
\end{align}
where $\mathbf{m^{\prime}}$ is a Bloch vector of $\mathcal{E}(\rho)$ and $m^{\prime}$ is the length of vector $\mathbf{m^{\prime}}$, and $\rho$ is a pure input state of a qubit channel $\mathcal{E}$, i.e., $\rho=\frac{1}{2}(I+\mathbf{m}\cdot\bf{\sigma})$ with $\mathbf{m}$ being unit real vector and $\mathbf{\sigma}=\{\sigma_{1},\sigma_{2},\sigma_{3}\}$ denoting the vector for Pauli matrices. $\mathbf{k}$ is a unit real vector standing for the reference basis $\{\frac{1}{2}(I+\mathbf{k}\cdot\mathbf{\sigma}), \frac{1}{2}(I-\mathbf{k}\cdot\mathbf{\sigma})\}$.

The decohering power of a channel is the maximum reduce of coherence caused by quantum channel acting on the maximally coherent states. The decohering power of a channel is defined as
\begin{align}
\mathcal{D}_{\mathbf{k}}(\mathcal{E})=\max_{\rho\in\mathcal{M}}\{C_{\mathbf{k}}(\rho)-C_{\mathbf{k}}(\mathcal{E}(\rho))\},
\end{align}
where $\mathcal{M}$ is the set of maximally coherent states.
For a quantum channel $\mathcal{E}$, decohering power can be evaluated through \cite{Mani2015}
\begin{align}
\mathcal{D}_{\mathbf{k}}(\mathcal{E})=1-\min_{\mathbf{m}} F_{\mathcal{E}}(\mathbf{m},\mathbf{k}).\label{DP}
\end{align}
The input state is an equatorial pure state with respect to the direction $\mathbf{k}$, i.e., $\mathbf{m}\bot\mathbf{k}$. In other words, the input state is of the form $|\phi\rangle=\frac{1}{\sqrt{2}}(|k+\rangle+e^{i \varphi}|k-\rangle)$, where $|k\pm\rangle$ are two eigenvectors of $\bf{\sigma}\cdot\mathbf{k}$.

Quantum correlation is the non-classical correlation beyond entanglement.
It's also an important resource in most quantum information processing tasks. One kind of quantum correlation measure called quantum discord, which was first proposed by Ollivier and Zurek \cite{Ollivier2001} and by Henderson and Vedral \cite{Henderson2001}, has received considerable attention. Calculating the quantum discord based on a numerical optimization procedure is a hard work. The difficulty of calculating the quantum discord led Daki\'{c}, et al. \cite{Dakic2010} to introduce a geometric quantum discord. The geometric quantum discord of a bipartite quantum state $\rho$ in Hilbert space $\mathcal{H}_{A}\otimes\mathcal{H}_{B}$ is defined \cite{Dakic2010,Luo2010} as
\begin{align}
D_{G}(\rho)=\min_{\rho_{c}\in\Omega_{0}}\|\rho-\rho_{c}\|_{2}^{2},
\end{align}
where $\|X\|_{2}=\sqrt{\text{Tr}(X^{\dag}X)}$ denotes the Hilbert-Schmidt norm and $\Omega_{0}$ is the set of zero-discord states (classical-quantum states, given by $\rho_{c}=\sum p_{k}|k\rangle \langle k|\otimes\rho_{k}$).
An arbitrary two-qubit state can be represented by
\begin{align}
\rho=\frac{1}{4}(I\otimes I+\sum_{i=1}^{3}x_{i}\sigma_{i}\otimes I+\sum_{i}^{3}I\otimes y_{i}\sigma_{i}+\sum_{i,j=1}^{3}t_{ij}\sigma_{i}\otimes \sigma_{j}),\label{eqstates}
\end{align}
with $x_{i}=\text{Tr}(\rho\sigma_{i}\otimes I),\ y_{i}=\text{Tr}(\rho I\otimes \sigma_{i}),\ t_{ij}=\text{Tr}(\rho \sigma_{i}\otimes \sigma_{j})$ being real parameters, and $\sigma_{i}$ being Pauli matrices. According to the above equation, geometric quantum discord can be calculated by \cite{Dakic2010,Luo2010}
\begin{align}
D_{G}(\rho)=\frac{1}{4}(\|\mathbf{x}\|^{2}+\|T\|^{2}_{2}-\lambda_{max}),\label{gqd1}
\end{align}
where $\mathbf{x}=(x_{1},x_{2},x_{3})^{t}$ is a column vector, $\|\mathbf{x}\|^{2}=\sum_{i}x_{i}^{2}$, $T=(t_{ij})$ is a matrix, and $\lambda_{max}$ is the largest eigenvalue of the matrix $\mathbf{x}\mathbf{x}^{t}+TT^{t}$.

Actually, there is an alternative concise calculating method of geometric quantum discord, reading \cite{Lu2010}
\begin{align}
D_{G}(\rho)=\frac{1}{4}(\sum_{i}\lambda_{i}^{2}-\max_{i}\lambda_{i}^{2}), \label{sgqd1}
\end{align}
with $\lambda_{i}$ being the singular values of the matrix $T^{\prime}=(\mathbf{x},T)$, a $3\times4$ matrix.

As we know, entanglement plays a central role in quantum information theory, which contains correlations that do not have a classical counterpart. The entanglement of formation is a monotonically increasing function of the concurrence \cite{Wootters1998}. For two-qubit mixed states, the concurrence is defined as
\begin{align}
C_{E}(\rho)=&\max\{0,2 \max\{\sqrt{\lambda_{1}},\sqrt{\lambda_{2}},\sqrt{\lambda_{3}},\sqrt{\lambda_{4}}\}\nonumber\\
&-\sqrt{\lambda_{1}}-\sqrt{\lambda_{2}}-\sqrt{\lambda_{3}}-\sqrt{\lambda_{4}}\},
\end{align}
where $\lambda_{i}$ are the eigenvalues of $\rho\tilde{\rho}$ in decreasing order, and $\tilde{\rho}=(\sigma_{2}\otimes\sigma_{2})\rho^{\ast}(\sigma_{2}\otimes\sigma_{2})$ with $\rho^{\ast}$ as the complex conjugated density matrix.

\section{Behaviors of quantum coherence in accelerated frame}\label{S4}
Quantum coherence can exhibit the most essential quantum feature in a single system. In this section, we study the behavior of quantum coherence for one party system in the Grassmann scalar field, which is very useful to study the general features of quantum correlation in fermionic fields \cite{Bruschi2010,Montero22011,Montero2012,Chang2012,Ramzan2014}. We suppose that Bob travels with a uniform acceleration. Then, the initial fermionic state of Bob can be described as
\begin{align}
|\phi^{+}\rangle=\cos(\theta)|0\rangle_{U}+\sin(\theta)|1^{+}\rangle_{U},\label{inst}
\end{align}
where $0\leq\theta\leq\frac{\pi}{4}$. When $\theta=\frac{\pi}{4}$, $|\phi^{+}\rangle$ become the maximally coherent state.

Suppose that Bob\rq s detector is only sensitive to particle. As is depicted in Fig. \ref{figure1}, since two regions I and II are causally disconnected, according to Eq. (\ref{U0}) and Eq. (\ref{U1}), by tracing over region II and the antiparticle in region I, the state can be obtained beyond the single-mode approximation
\begin{align}
\rho^{\phi^{+}}_{B_{I}^{+}}=&(-q_{R}^{2}\cos^{2}\theta\cos^{2}r+\cos^{2}r)|0\rangle\langle 0|\nonumber\\
&+\frac{q_{R}}{2}\sin 2\theta\cos r|0\rangle\langle 1|+\frac{q_{R}}{2}\sin 2\theta\cos r|1\rangle\langle 0|\nonumber\\
&+(q_{R}^{2}\cos^{2}\theta\cos^{2}r+\sin^{2}r)|1\rangle\langle 1|.
\end{align}
According to Eq. (\ref{l1}), $l_1$ norm of coherence can be obtained
\begin{align}
C_{l_{1}}(\rho^{\phi^{+}}_{B_{I}^{+}})=q_{R}\sin2\theta\cos r.
\end{align}
We are particularly interested in whether the coherence of the final state can be frozen under some initial or mode approximation conditions. Such conditions can be obtained by differentiating $C_{l_{1}}(\rho^{\phi^{+}}_{B_{I}^{+}})$ with respect to the acceleration parameter $r$. That is, the coherence is unaffected by acceleration if the differential is zero. the $r$ derivative of the $l_1$ norm can be calculated as follow
\begin{align}
\partial_{r}C_{l_{1}}(\rho^{\phi^{+}}_{B_{I}^{+}})=-q_{R}\sin2\theta\sin r,
\end{align}
which equals zero only for $\sin2\theta=0$ or $q_{R}=0$. Thus the freezing conditions for $l_1$ norm of coherence are that initial state should be an incoherence state or $q_{R}=0$, i.e., $q_{L}=1$. Actually, when $q_{R}=0$, $\rho^{\phi^{+}}_{B_{I}^{+}}=\cos^{2}r|0\rangle\langle 0|+\sin^{2}r|1\rangle\langle 1|$, which is an incoherence state.

Now assume that Bob\rq s detector is only sensitive to antiparticle. By tracing over region II and the particle in region I, the state is given as
\begin{align}
\rho^{\phi^{+}}_{B_{I}^{-}}=&(q_{L}^{2}\cos^{2}\theta\sin^{2}r+\cos^{2}r)|0\rangle\langle 0|\nonumber\\
&-\frac{q_{L}}{2}\sin 2\theta\sin r|0\rangle\langle 1|-\frac{q_{L}}{2}\sin 2\theta\sin r|1\rangle\langle 0|\nonumber\\
&+(-q_{L}^{2}\cos^{2}\theta\sin^{2}r+\sin^{2}r)|1\rangle\langle 1|.
\end{align}
$l_1$ norm of coherence is calculated as follow
\begin{align}
C_{l_{1}}(\rho^{\phi^{+}}_{B_{I}^{-}})=q_{L}\sin2\theta\sin r.
\end{align}
Analogously, by taking the $r$ derivative of $C_{l_{1}}(\rho^{\phi^{+}}_{B_{I}^{-}})$, we can obtain the coherence freezing conditions that initial state should be an incoherence state and $q_{L}=0 (q_{R}=1)$ corresponding to the single-mode approximation. When $q_{R}=0$, we get the same incoherent state $\cos^{2}r|0\rangle\langle 0|+\sin^{2}r|1\rangle\langle 1|$.

\begin{figure}
\centering
\subfigure[]{\includegraphics[height=3.8cm,width=3.8cm]{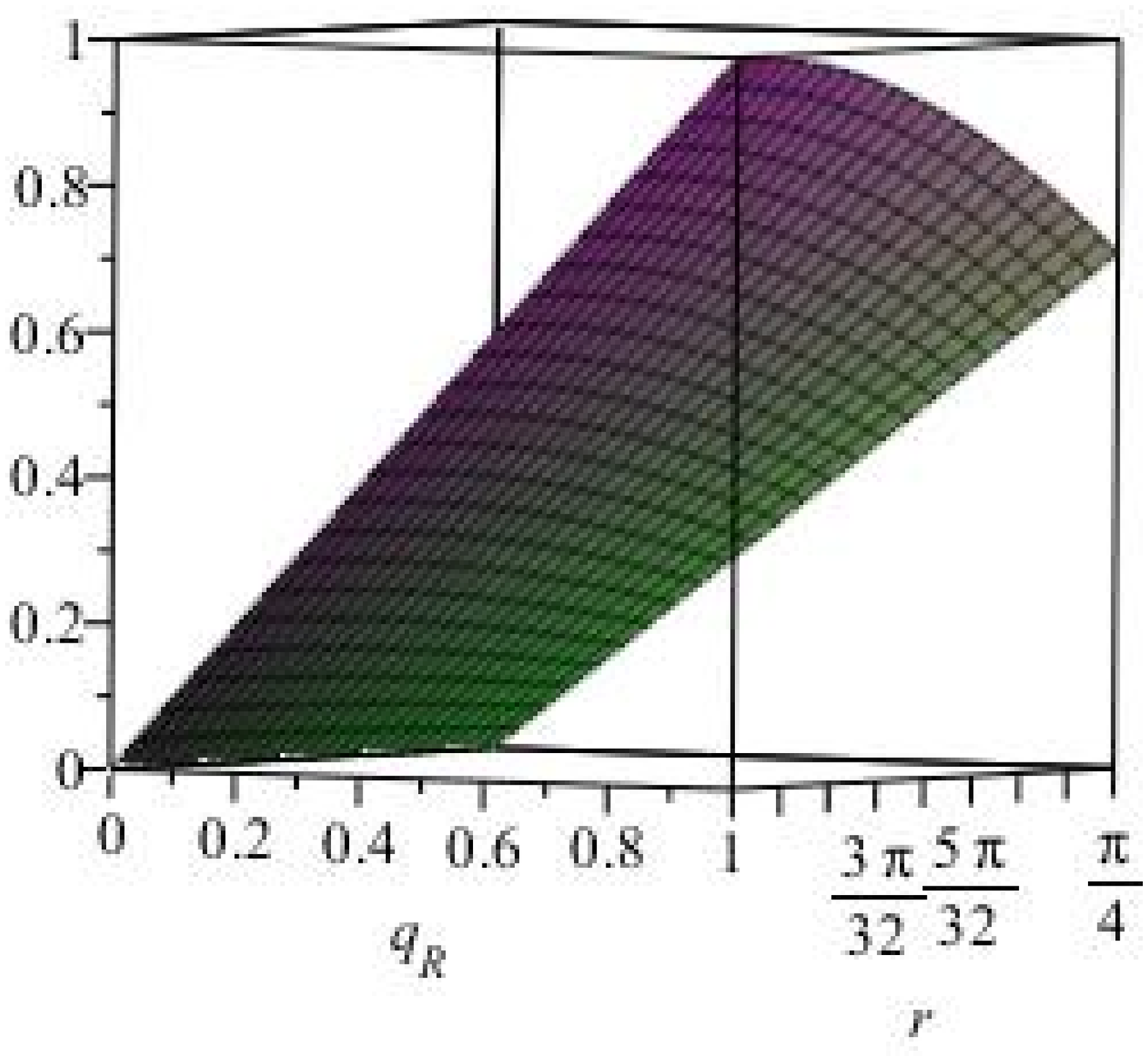}}
\subfigure[]{\includegraphics[height=3.8cm,width=3.8cm]{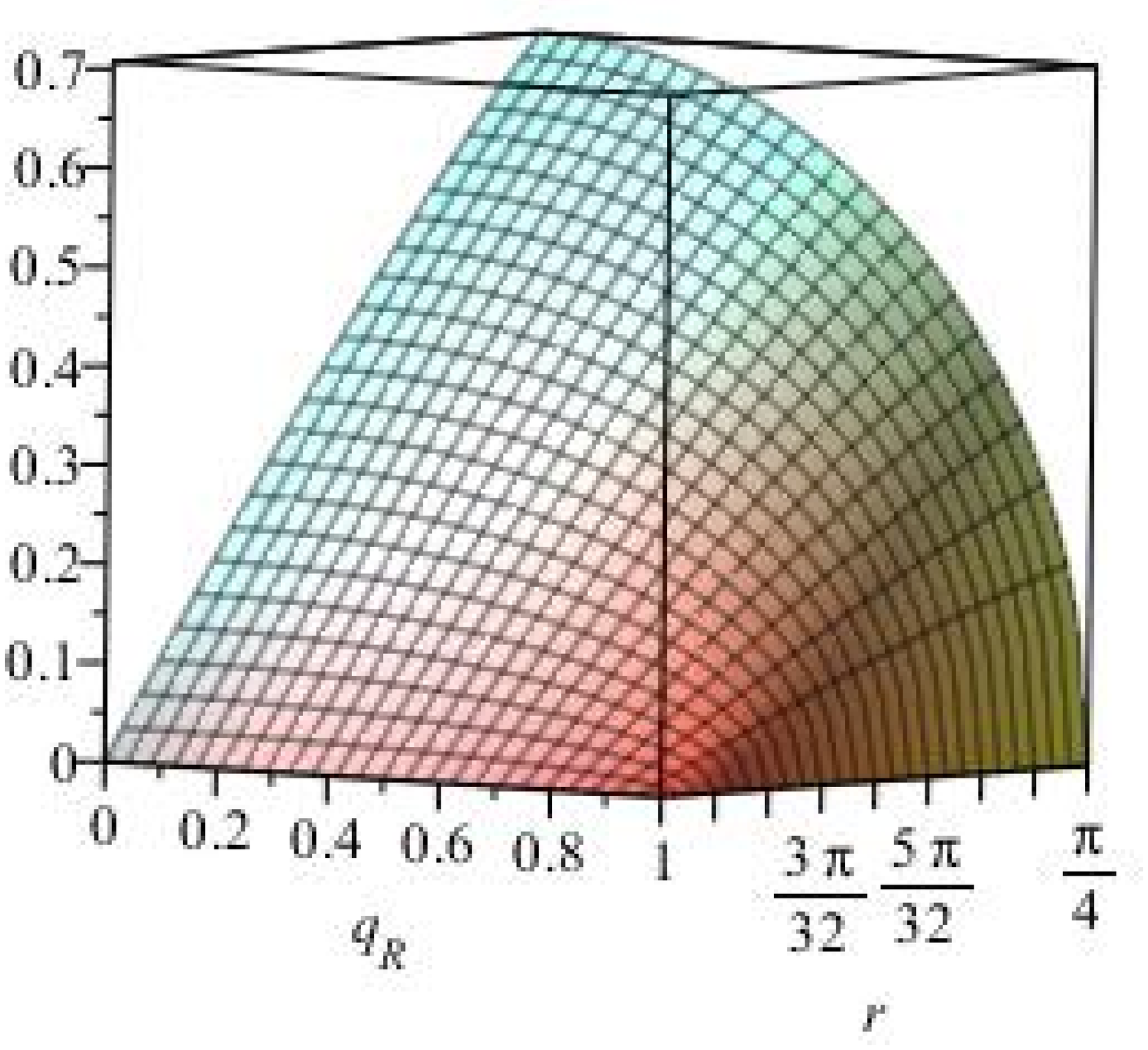}}
\caption{\label{figureb}(a) $l_1$ norm of coherence for particle in region I and (b) $l_1$ norm of coherence for antiparticle in region I as a function of acceleration $r$ and parameter $q_{R}$ with $\theta=\frac{\pi}{4}$.}
\end{figure}
Fig. \ref{figureb} depicts the $l_1$ norm of coherence for particle state $\rho^{\phi^{+}}_{B_{I}^{+}}$ and antiparticle $\rho^{\phi^{+}}_{B_{I}^{-}}$ in region I. From Fig. \ref{figureb}, we can observe that the coherence of particle sector grows monotonically as the growth of the parameter $q_{R}$ and degrades as the acceleration $r$ increases. However, the coherence in antiparticle sector shows reverse variation. Therefore, there is coherence transfer between particle sector and antiparticle sector, but the coherence lost in particle sector is not completely compensated by the coherence creation of antiparticle sector. When $q_{R}=1$ and $r=0$, the coherence of particle sector reaches the maximal value $1$. When $q_{R}=0$ (or $q_{L}=1$) and $r=\frac{\pi}{4}$ (infinite-acceleration limit), the maximal coherence in antiparticle sector is $\frac{\sqrt{2}}{2}$. It is noted that when $q_{R}=0$ and $r=0$, particle and antiparticle sector have vanishing coherence. The equal critical points of coherence in two sectors can be localized by the following function
\begin{align}
\tan r=\frac{q_{R}}{q_{L}}=\frac{q_{R}}{\sqrt{1-q_{R}^{2}}},
\end{align}
which means when acceleration $r$ and parameter $q_{R}$ satisfy the above relation, the coherence in particle sector equals the lost coherence to antiparticle sector.

As discussed above, coherence decays in particle sector and grows in anitparticle sector with increase of acceleration. Hence now, we investigate the cohering power and decohering power in accelerated frame. For simplicity, in the following discussion, we focus on our analysis with respect to the computational basis ($\mathbf{z}$ basis), namely, $\mathbf{k}=(0,0,1)^{t}$.

We fist check the cohering power and decohering power in particle sector of region I. According to Eq. (\ref{CP}), the input pure states are represented as $\rho=\frac{1}{2}(I+\sigma_{3})$ and $\rho=\frac{1}{2}(I-\sigma_{3})$, respectively. Accordingly, the output states of Unruh channel are $\mathcal{E}_{U}(\rho)=\cos^{2}r|0\rangle\langle 0|+\sin^{2}r|1\rangle\langle 1|$ and $\mathcal{E}_{U}(\rho)=q_{L}^{2}\cos^{2}r|0\rangle\langle 0|+(q_{L}^{2}\sin^{2}r+q_{R}^{2})|1\rangle\langle 1|$, respectively, both of which are incoherent states. By definition, we have
\begin{align}
\mathcal{C}_{\mathbf{z}}(\mathcal{E}_{U})=0.
\end{align}

Next, let us investigate the decohering power of Unruh channel in particle sector. We can suppose that the input pure state is $|\phi\rangle=\frac{1}{\sqrt{2}}(|0\rangle+e^{i \varphi}|1\rangle)$ . After through the Unruh channel, we have $\mathcal{E}_{U}(|\phi\rangle\langle\phi|)=\frac{1}{2}[I+q_{R}\cos\varphi\cos r\sigma_{1}+q_{R}\sin\varphi\cos r\sigma_{2}+(q_{L}^{2}\cos^{2}r-\sin^{2}r)\sigma_{3}]$. By using Eq. (\ref{FC}) and Eq. (\ref{DP}), doing simplifications, we obtain
\begin{align}
\mathcal{D}_{\mathbf{z}}(\mathcal{E}_{U})&=1-\min_{\varphi}\{(1-\sqrt{1-mv})[1-\frac{(q_{L}^{2}\cos^{2} r-\sin^{2} r)^{2}}{mv}]\}\nonumber\\
&=1-(1-\sqrt{1-mv})[1-\frac{(q_{L}^{2}\cos^{2} r-\sin^{2} r)^{2}}{mv}],
\end{align}
where $mv=q_{R}^{2}\cos^{2} r+(q_{L}^{2}\cos^{2} r-\sin^{2} r)^{2}$.

We also easily find that the cohering power vanishes in antiparticle sector. With similar analysis, the decohering power in antiparticle sector is given by
\begin{align}
\mathcal{D}_{\mathbf{z}}(\mathcal{E}_{U})&=1-(1-\sqrt{1-mv})[1-\frac{(q_{L}^{2}\sin^{2} r+\cos 2r)^{2}}{mv}],
\end{align}
where $mv=q_{L}^{2}\sin^{2} r+(q_{L}^{2}\sin^{2} r+\cos 2r)^{2}$.

\begin{figure}
\centering
\subfigure[]{\includegraphics[height=3.8cm,width=3.8cm]{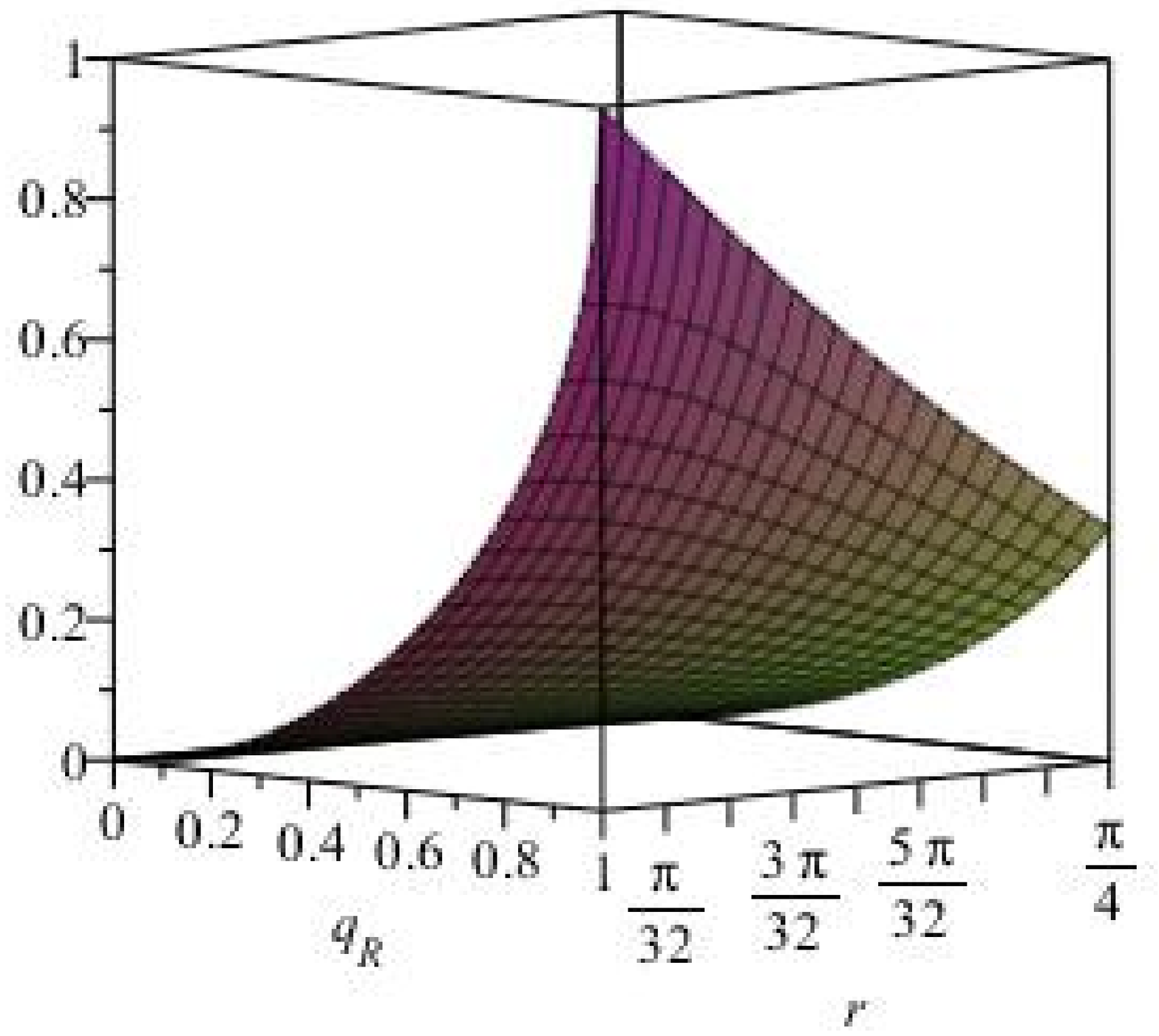}}
\subfigure[]{\includegraphics[height=3.8cm,width=3.8cm]{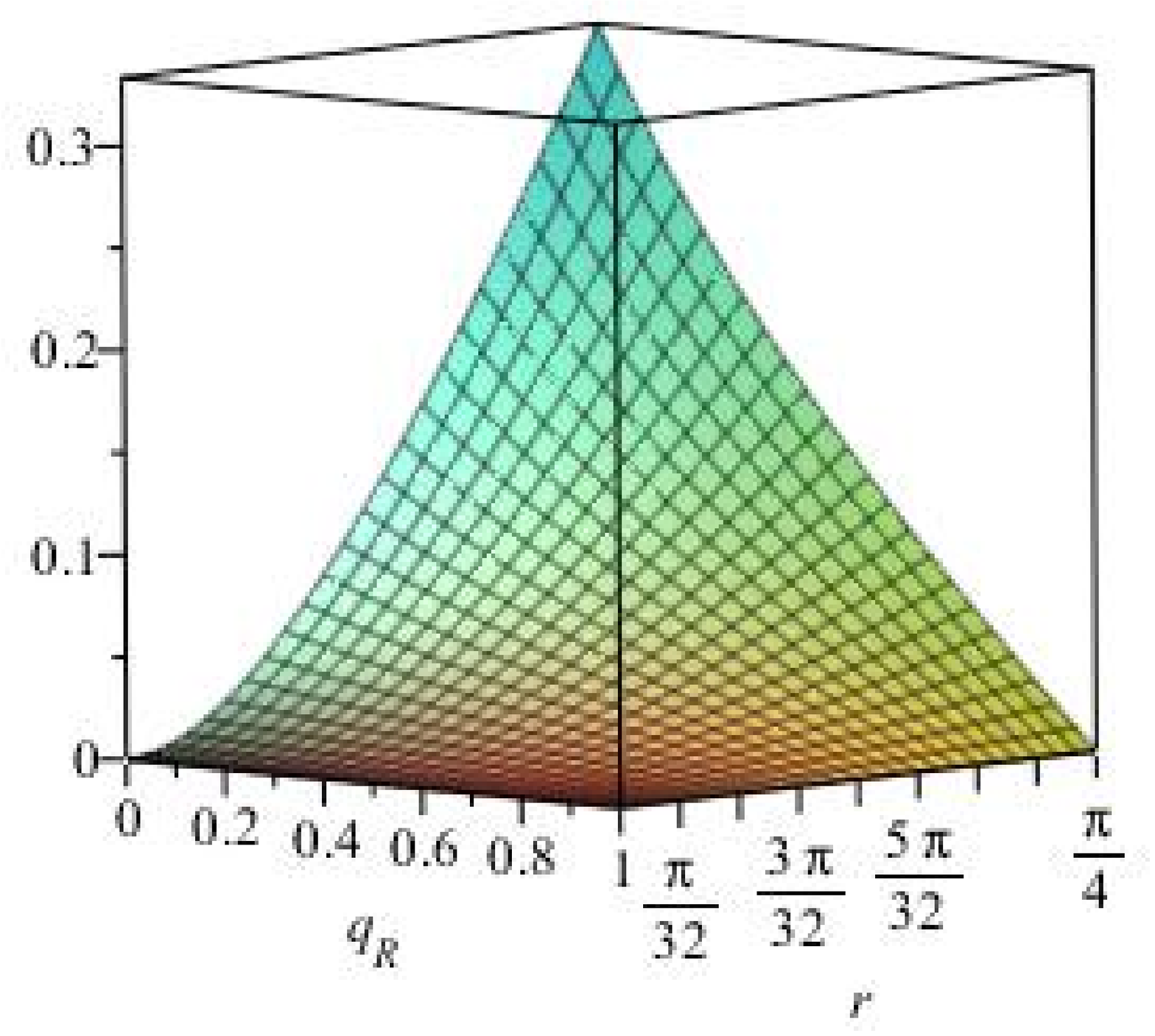}}
\caption{\label{figuredp} Decohering power in particle sector (a) and antiparticle sector (b) as a function of acceleration $r$ and parameter $q_{R}$.}
\end{figure}
Fig. \ref{figuredp} shows the decohering power variation with acceleration $r$ and parameter $q_{R}$. It is seen that decohering power of Unruh channel in particle sector increases monotonically with the growth of parameter $q_{R}$, and decreases with the acceleration $r$ increasing. However, in antiparticle sector, the decohering power has completely opposite changes.

Now, we analyse the dynamics of fermionic coherence in noninertial frames. Suppose that Bob\rq s detector can not distinguish particle and antiparticle and the initial state is the maximally coherent state (taking $\theta=\frac{\pi}{4}$ in Eq. (\ref{inst})).
After tracing out the region II, the state in region I is described as
\begin{align}
\rho^{\phi^{+}}_{B_{I}}=&\frac{1}{2}\cos^{2}r(q_{L}^{2}+\cos^{2}r)|00\rangle\langle00|-\frac{q_{L}}{2}\cos^{2}r\sin r(|00\rangle\langle01|+|01\rangle\langle00|)\nonumber\\
&+\frac{q_{R}}{2}\cos^{3}r(|00\rangle\langle10|+|10\rangle\langle00|)-\frac{q_{L}q_{R}}{4}\sin 2r(|00\rangle\langle11|+|11\rangle\langle00|)\nonumber\\
&+\frac{1}{2}\cos^{2}r\sin^{2}r|01\rangle\langle01|+\frac{q_{R}}{2}\cos{2}r\sin^{2}r(|01\rangle\langle11|+|11\rangle\langle01|)\nonumber\\
&-(\frac{q_{L}^{2}}{2}\cos 2r+\frac{1}{2}\cos^{4}r-\cos^{2}r)|10\rangle\langle10|-\frac{q_{L}}{2}\sin^{3}r(|10\rangle\langle11|+|11\rangle\langle10|)\nonumber\\
&+\frac{1}{2}\sin^{2}r(\sin^{2}r+q_{R}^{2})|11\rangle\langle11|.
\end{align}
The density matrix of region II is obtained by tracing over region I:
\begin{align}
\rho^{\phi^{+}}_{B_{II}}=&\frac{1}{2}\cos^{2}r(q_{R}^{2}+\cos^{2}r)|00\rangle\langle00|+\frac{q_{R}}{2}\cos^{2}r\sin r(|00\rangle\langle10|+|10\rangle\langle00|)\nonumber\\
&+\frac{q_{L}}{2}\cos^{3}r(|00\rangle\langle01|+|01\rangle\langle00|)+\frac{q_{L}q_{R}}{4}\sin 2r(|00\rangle\langle11|+|11\rangle\langle00|)\nonumber\\
&+(\frac{q_{L}^{2}}{2}\cos 2r-\frac{1}{2}\cos^{4}r+\frac{1}{2})|01\rangle\langle01|+\frac{q_{R}}{2}\sin^{3}r(|01\rangle\langle11|+|11\rangle\langle01|)\nonumber\\
&+\frac{1}{2}\cos^{2}r\sin^{2}r|10\rangle\langle10|+\frac{q_{L}}{2}\cos r\sin^{2} r(|10\rangle\langle11|+|11\rangle\langle10|)\nonumber\\
&+\frac{1}{2}\sin^{2}r(\sin^{2}r+q_{L}^{2})|11\rangle\langle11|.
\end{align}
According to Eq. (\ref{l1}), $l_1$ norm of coherence in region I and region II, respectively are
\begin{align}
&C_{l_{1}}(\rho^{\phi^{+}}_{B_{I}})=q_{L}q_{R}\cos r\sin r+q_{L}\sin r+q_{R}\cos r,\\
&C_{l_{1}}(\rho^{\phi^{+}}_{B_{II}})=q_{L}q_{R}\cos r\sin r+q_{L}\cos r+q_{R}\sin r.
\end{align}
From the above equations, it is easily analysed that when $q_{L}=q_{R}$ or $\sin r=\cos r$, that is $q_{L}=q_{R}=\frac{\sqrt{2}}{2}$ or $r=\frac{\pi}{4}$ (infinite acceleration), $C_{l_{1}}(\rho^{\phi^{+}}_{B_{I}})=C_{l_{1}}(\rho^{\phi^{+}}_{B_{II}})$.
\begin{figure}
\centering
\subfigure[]{\includegraphics[height=3.8cm,width=3.8cm]{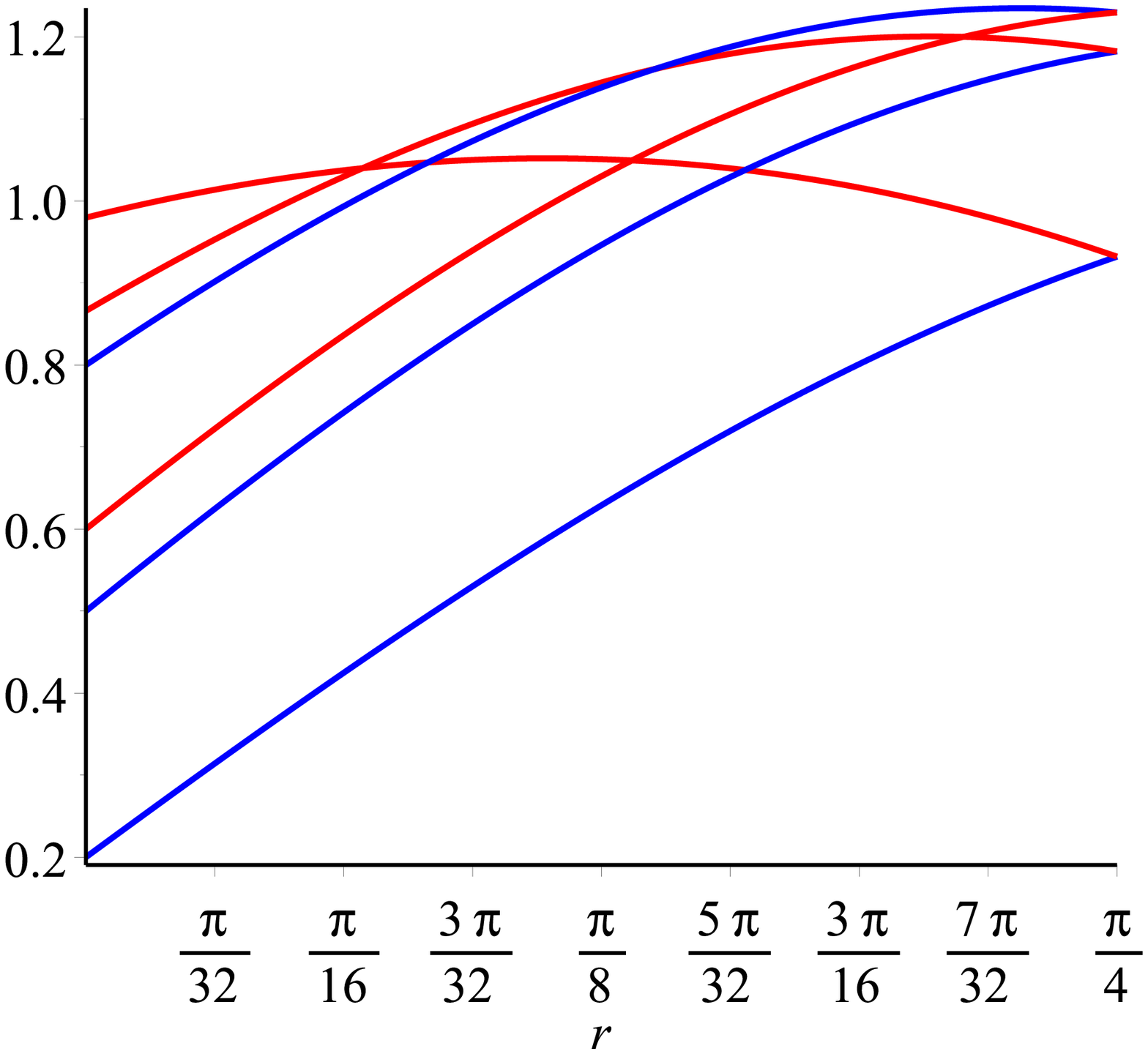}}
\subfigure[]{\includegraphics[height=3.8cm,width=3.8cm]{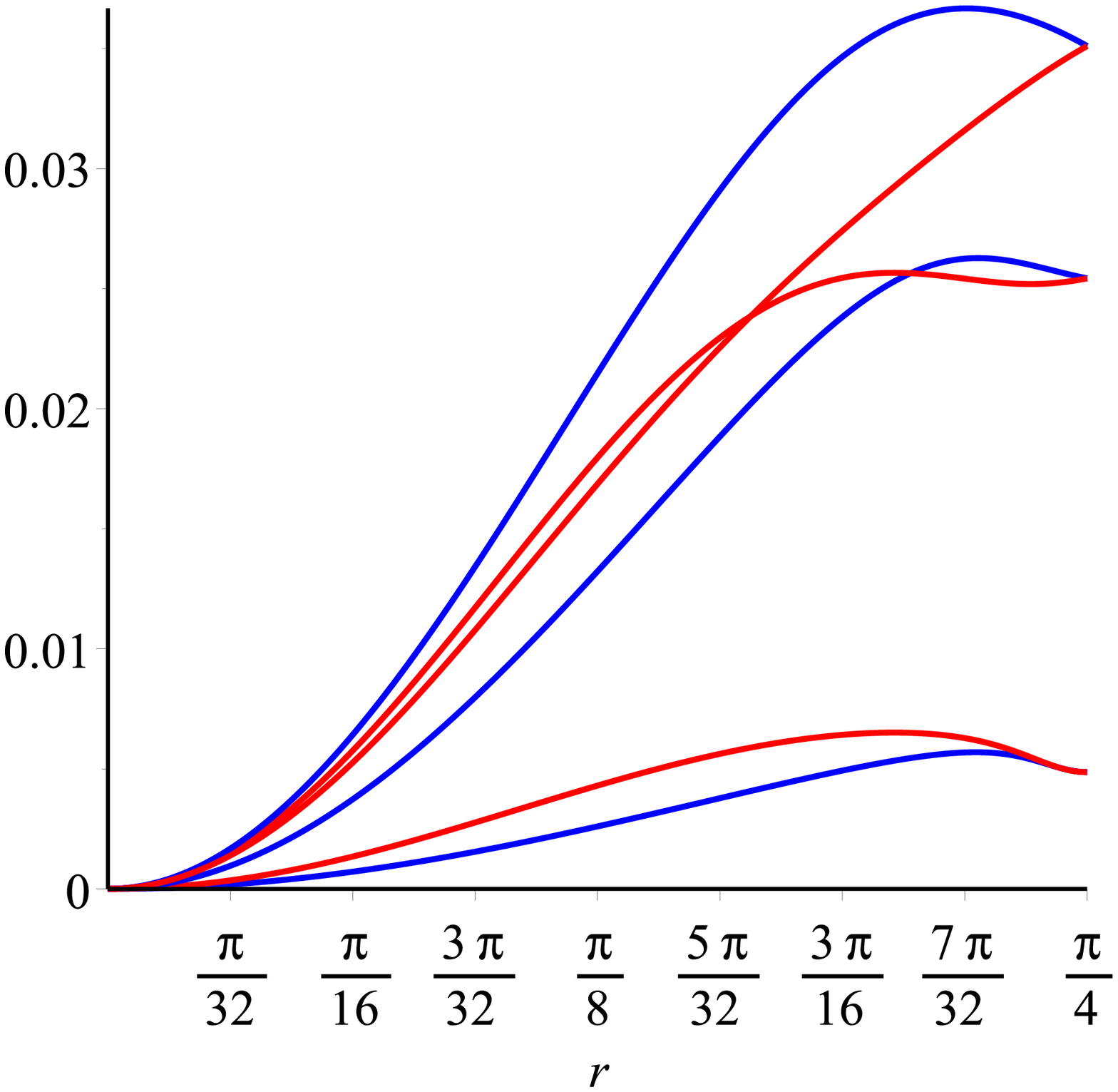}}
\subfigure[]{\includegraphics[height=3.8cm,width=3.8cm]{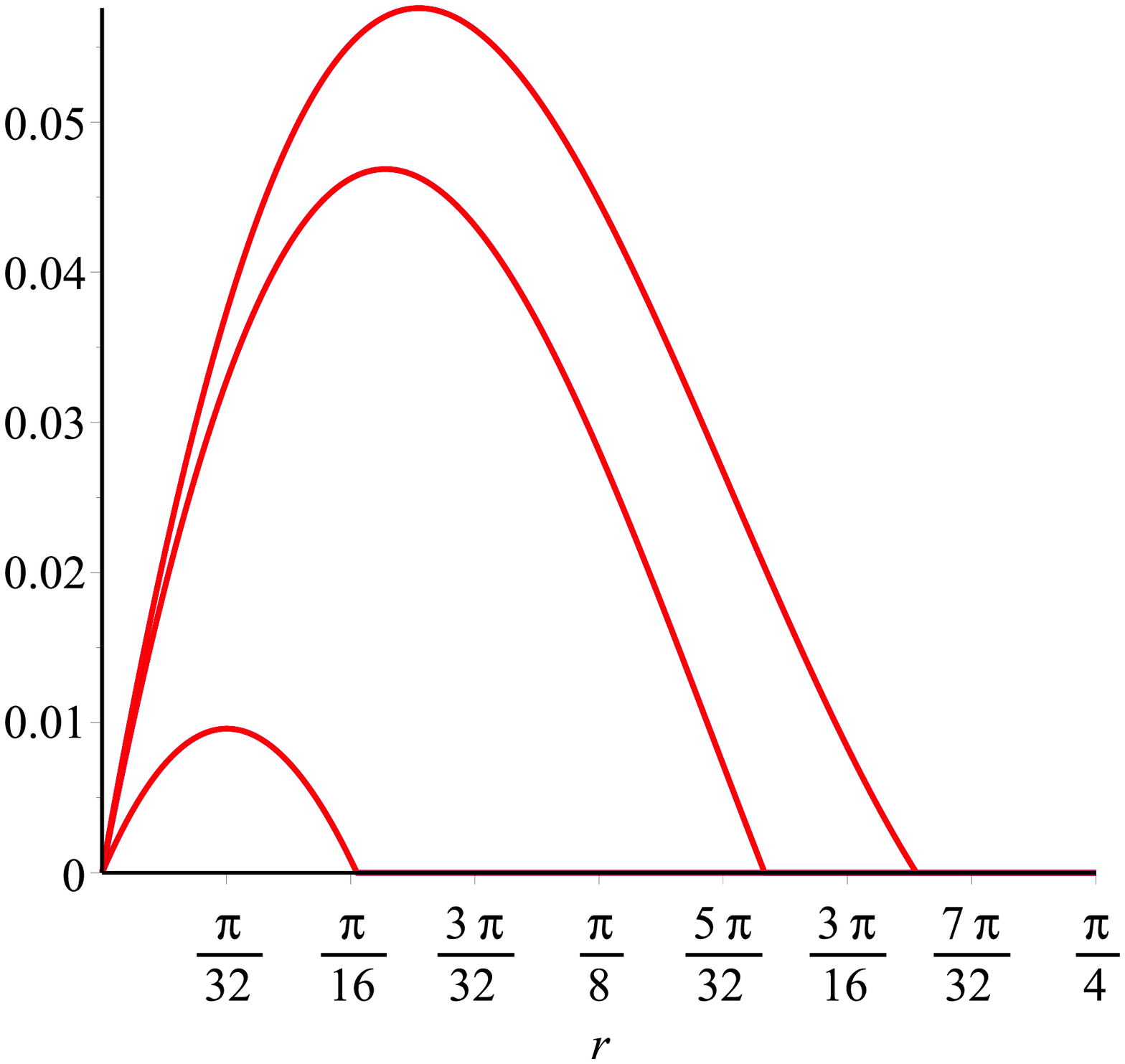}}
\caption{\label{figurec1}(a) $C_{l_{1}}(\rho^{\phi^{+}}_{B_{I}})$ (blue) and $C_{l_{1}}(\rho^{\phi^{+}}_{B_{II}})$ (red), (b) $D_{G}(\rho^{\phi^{+}}_{B_{I}})$ (blue) and $D_{G}(\rho^{\phi^{+}}_{B_{II}})$ (red), and (c) $C_{E}(\rho^{\phi^{+}}_{B_{I}})$ as a function of acceleration $r$ at $q_{R} = 0.2, 0.5$ and 0.8, respectively. }
\end{figure}
As it can be seen from Fig. \ref{figurec1}, the $l_1$ norm of coherence of $\rho^{\phi^{+}}_{B_{I}}$  and $\rho^{\phi^{+}}_{B_{II}}$ coincides at infinite acceleration, so does geometric quantum discord. We show that the behavior of quantum coherence and geometric quantum discord allows one to obtain physical results, i.e., they converge in the infinite acceleration \cite{Martin2011,Chang2012}.  This implies that the fermionic coherence and fermionic discord become independent of $q_{R}$ at infinite acceleration limit. Interestingly, $\rho^{\phi^{+}}_{B_{I}}$  and $\rho^{\phi^{+}}_{B_{II}}$ have the same entanglement behavior. It is obvious that entanglement increases and then decreases with the acceleration increasing, and has sudden death, which is in distinct contrast with the behaviors of quantum coherence and geometric quantum discord. The behaviors of this resource measures are very different from the behaviors of quantum discord and entanglement investigated in other papers \cite{Bruschi2010,Martin2011,Chang2012,Ramzan2014}.

Then, we consider the initial fermionic state
\begin{align}
|\phi^{-}\rangle=\cos(\theta)|0\rangle_{U}+\sin(\theta)|1^{-}\rangle_{U}.\label{inst1}
\end{align}
we trace over region II and the antiparticle in region I, and obtain
\begin{align}
\rho^{\phi^{-}}_{B_{I}^{+}}=&\cos^{2}r(1-q_{L}^{2}\cos^{2}\theta)|0\rangle\langle 0|\nonumber\\
&+\frac{q_{R}}{2}\sin 2\theta\sin r|0\rangle\langle 1|+\frac{q_{R}}{2}\sin 2\theta\sin r|1\rangle\langle 0|\nonumber\\
&+[q_{L}^{2}\cos^{2}\theta\sin^{2}r-\sin^{2}\theta(\cos^{2}r+1)]|1\rangle\langle 1|.
\end{align}
According to Eq. (\ref{l1}), $l_1$ norm of coherence of $\rho^{\phi^{-}}_{B_{I}^{+}}$ can be obtained
\begin{align}
C_{l_{1}}(\rho^{\phi^{-}}_{B_{I}^{+}})=q_{R}\sin2\theta\sin r.
\end{align}
We are also interested in the freezing condition of $l_1$ norm. By differentiating $C_{l_{1}}(\rho^{\phi^{-}}_{B_{I}^{+}})$ with respect to the acceleration $r$, we get
\begin{align}
\partial_{r}C_{l_{1}}(\rho^{\phi^{-}}_{B_{I}^{+}})=q_{R}\sin2\theta\cos r.
\end{align}
We easily gain the freezing conditions for $l_1$ norm of coherence that the initial state is an incoherent state or $q_{R}=0$, i.e., $q_{L}=1$.

The density matrix of  antiparticle state is obtained by tracing over region II and the particle in region I
\begin{align}
\rho^{\phi^{-}}_{B_{I}^{-}}=&\cos^{2}r(1-q_{L}^{2}\cos^{2}\theta)|0\rangle\langle 0|\nonumber\\
&+\frac{q_{L}}{2}\sin 2\theta\sin r|0\rangle\langle 1|+\frac{q_{L}}{2}\sin 2\theta\sin r|1\rangle\langle 0|\nonumber\\
&+(q_{L}^{2}\cos^{2}\theta\cos^{2}r-\sin^{2}r)|1\rangle\langle 1|.
\end{align}
The $l_1$ norm of coherence of $\rho^{\phi^{-}}_{B_{I}^{-}}$ is gained
\begin{align}
C_{l_{1}}(\rho^{\phi^{-}}_{B_{I}^{-}})=q_{L}\sin2\theta\cos r.
\end{align}
Similarly, by taking the $r$ derivative of $C_{l_{1}}(\rho^{\phi^{-}}_{B_{I}^{-}})$, we can derive the coherence freezing conditions that initial state should be an incoherent state and $q_{L}= 0$ $(q_{R}= 1)$.
\begin{figure}
\centering
\subfigure[]{\includegraphics[height=3.8cm,width=3.8cm]{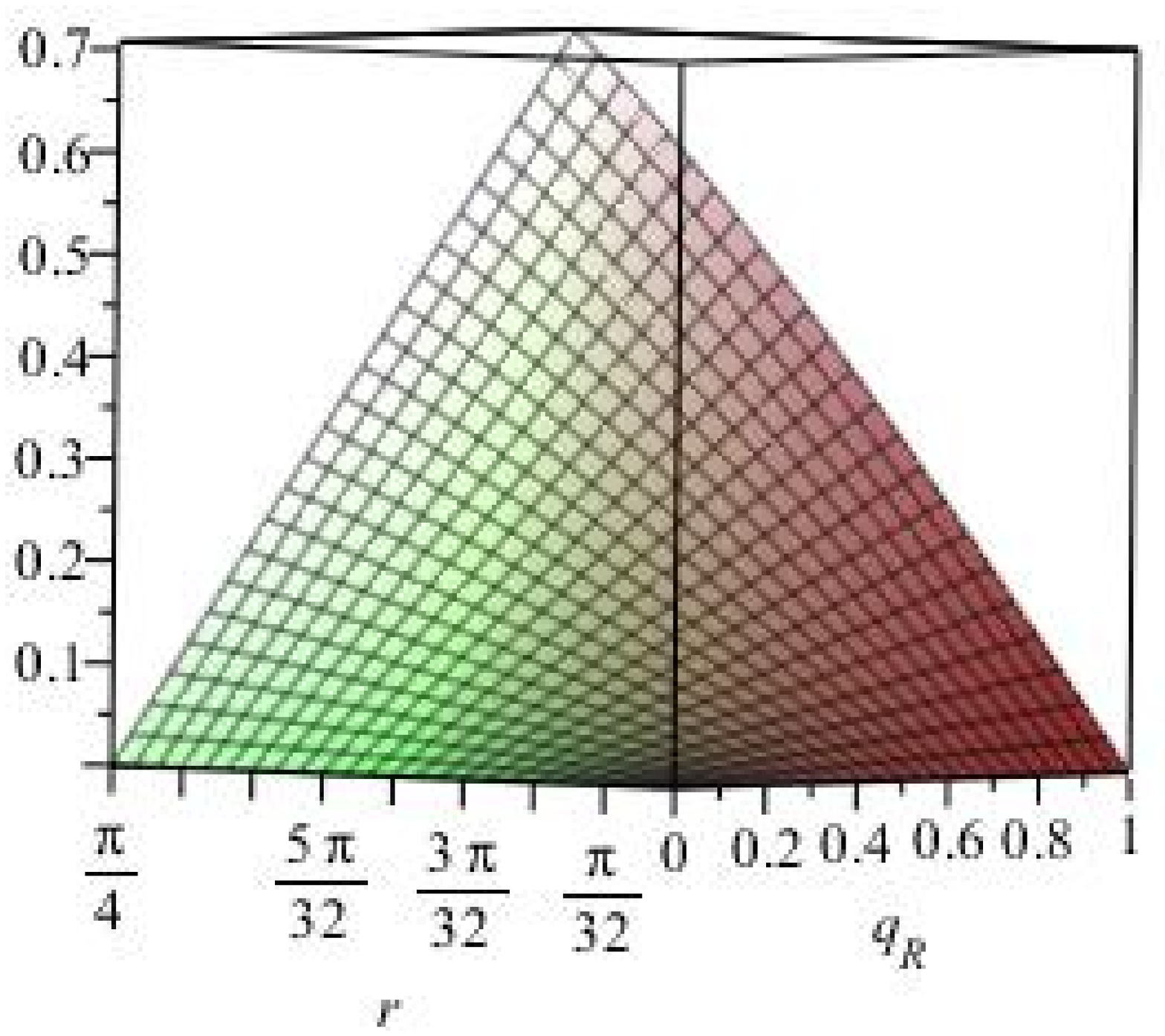}}
\subfigure[]{\includegraphics[height=3.8cm,width=3.8cm]{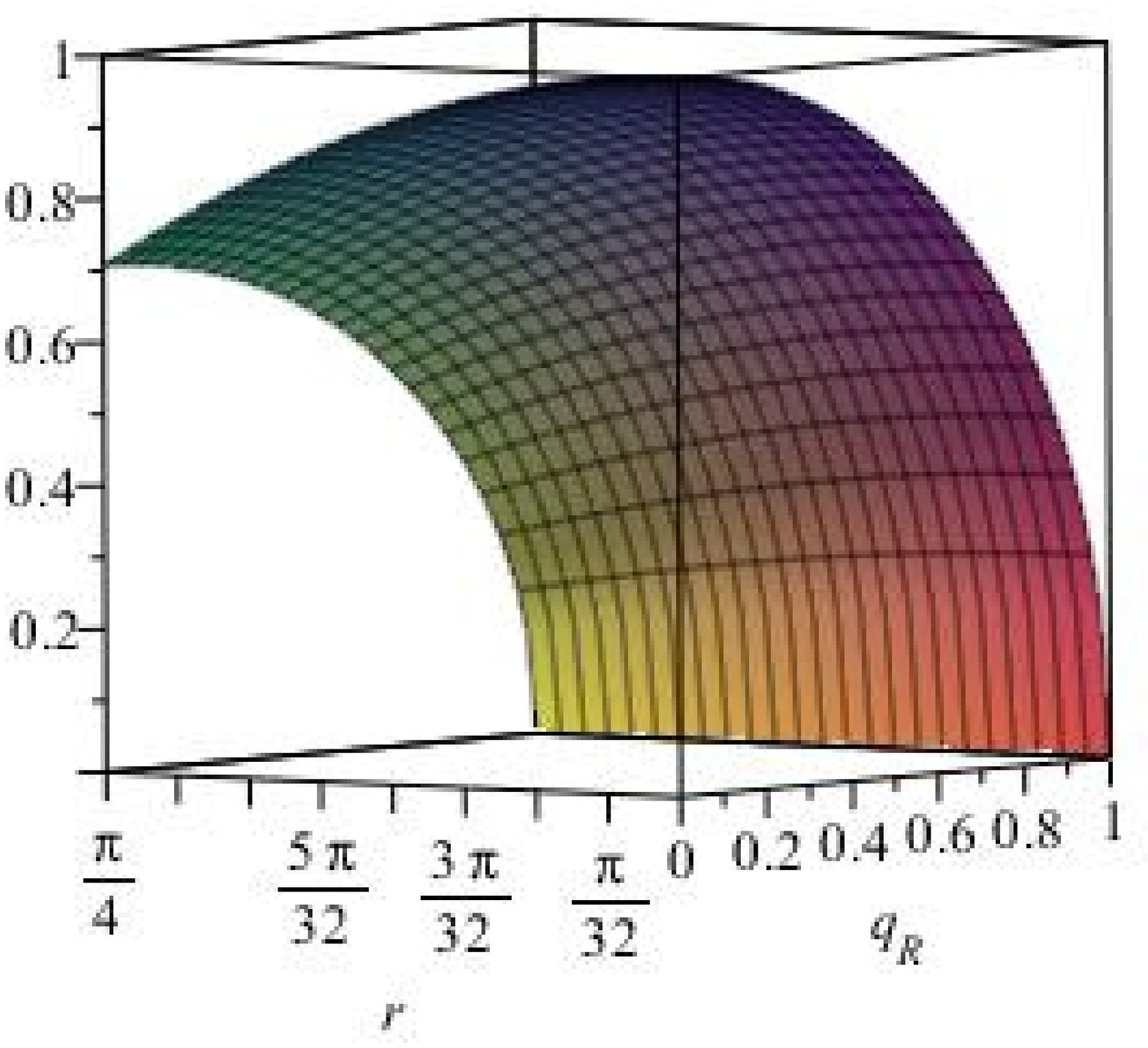}}
\caption{\label{figured}(a) $l_1$ norm of coherence for particle in region I and (b) $l_1$ norm of coherence for antiparticle in region I as a function of acceleration $r$ and parameter $q_{R}$ with $\theta=\frac{\pi}{4}$.}
\end{figure}

In Fig. \ref{figured}, we plot the $l_1$ norm of coherence for particle state $\rho^{\phi^{-}}_{B_{I}^{+}}$ and antiparticle $\rho^{\phi^{-}}_{B_{I}^{-}}$ in region I. From Fig. \ref{figured}, we can see that the coherence of particle sector grows monotonically as the growth of the parameter $q_{R}$ and rises as the acceleration $r$ increases, and the coherence in antiparticle sector is in the opposite trend simultaneously, which is different from the case of state $|\phi^{+}\rangle$. Analogously, there is coherence transfer between particle sector and antiparticle sector. When $q_{R}=1$ and $r=\frac{\pi}{4}$ (infinite-acceleration limit), the coherence of particle sector reaches the maximal value $\frac{\sqrt{2}}{2}$. When $q_{R}=0$ (or $q_{L}=1$) and $r=0$ ($r=\frac{\pi}{4}$), the maximal coherence in antiparticle sector are $1$ ($\frac{\sqrt{2}}{2}$). It is noted that when $q_{R}=1$ and $r=0$, the coherence of particle and antiparticle sector are zero. The equal critical points of coherence in two sectors can be determined by the following equation
\begin{align}
\tan r=\frac{q_{L}}{q_{R}}=\frac{\sqrt{1-q_{R}^{2}}}{q_{R}}.
\end{align}
Above equation implies when acceleration $r$ and parameter $q_{R}$ meet above relation, the coherence in particle sector equals the lost coherence.

With similar analysis, we can get cohering power $\mathcal{C}_{\mathbf{z}}(\mathcal{E}_{U})=0$ in Unruh channel with respect to  Eq. (\ref{U0}) and Eq. (\ref{U2}) in particle sector.  The decohering power in particle sector is given as
\begin{align}
\mathcal{D}_{\mathbf{z}}(\mathcal{E}_{U})=1-(1-\sqrt{1-mv})[1-\frac{(-q_{L}^{2}\sin^{2} r+\cos^{2}r)^{2}}{mv}],
\end{align}
where $mv=q_{R}^{2}\sin^{2} r+(-q_{L}^{2}\sin^{2} r+\cos^{2}r)^{2}$.
The decohering power in particle sector is given by
\begin{align}
\mathcal{D}_{\mathbf{z}}(\mathcal{E}_{U})=1-(1-\sqrt{1-mv})[1-\frac{(-q_{L}^{2}\cos^{2} r+\cos2r)^{2}}{mv}],
\end{align}
where $mv=q_{L}^{2}\cos^{2} r+(-q_{L}^{2}\cos^{2} r+\cos2r)^{2}$.
\begin{figure}
\centering
\subfigure[]{\includegraphics[height=3.8cm,width=3.8cm]{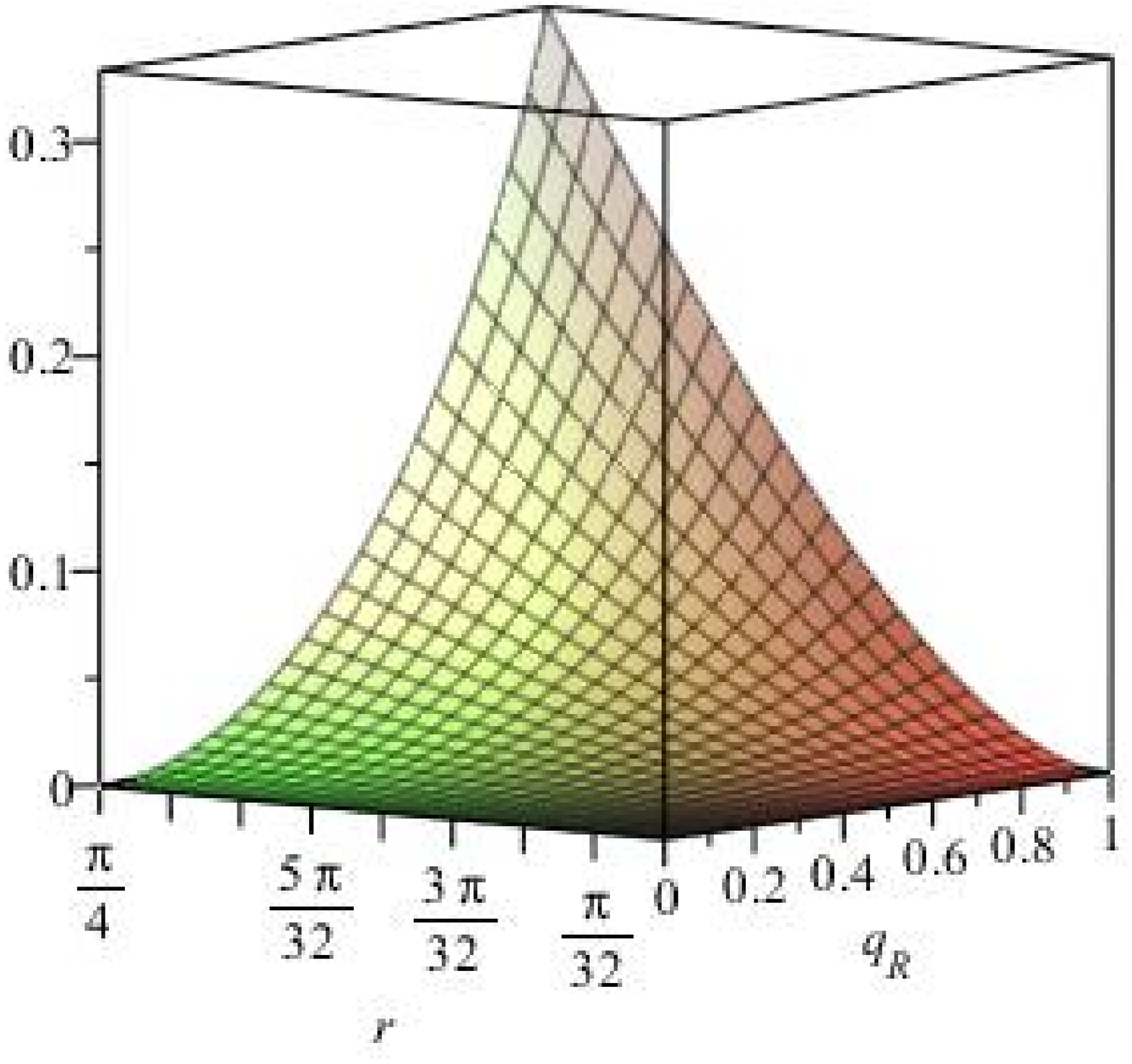}}
\subfigure[]{\includegraphics[height=3.8cm,width=3.8cm]{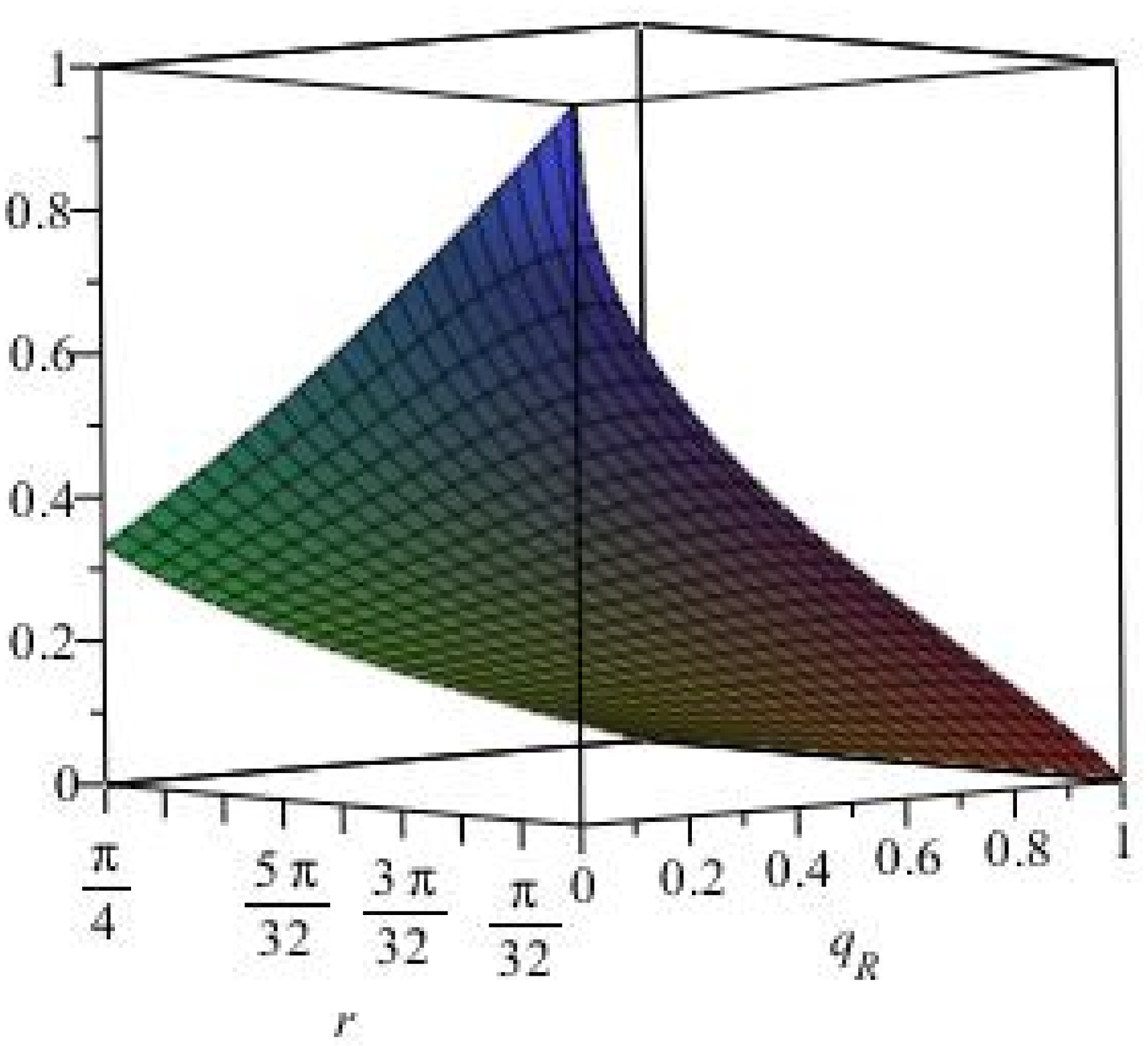}}
\caption{\label{figuredp2} Decohering power in particle sector (a) and antiparticle sector (b) as a function of acceleration $r$ and parameter $q_{R}$.}
\end{figure}

Fig. \ref{figuredp2} shows the decohering power variation with acceleration $r$ and parameter $q_{R}$. It is seen that decohering power of Unruh channel in particle sector increases with the growth of parameter $q_{R}$, and grows up with the increasing acceleration. However, in antiparticle sector, the decohering power has completely opposite trends.

Interestingly, we find the behaviors of quantum coherence, geometric quantum discord and entanglement in Unruh channel with respect to  Eq. (\ref{U0}) and Eq. (\ref{U2}) resemble the behaviors in Unruh channel with respect to  Eq. (\ref{U0}) and Eq. (\ref{U1}), so here we omit the discussion of that case.
\section{Conclusion}\label{S5}
In this paper, we have studied the quantum coherence behavior of fermionic system in noninertial frame. We find that the freezing conditions are that the initial state
is prepared as an incoherence state, or that the Unruh mode is single-mode approximation. We have analyzed the quantum coherence redistribution between particle and antiparticle modes of Grassman scalars field. Meanwhile, we discuss the cohering power and decohering power of Unruh channel. We discover that cohering power is trivial, but decohering power is dependent of the choice of Unruh mode and acceleration, and different regions have different variations. Besides, we compare and analyse the dynamics of quantum coherence, geometric quantum discord and entanglement. It is shown that these measures converge at infinite acceleration limit, which means that they become independent of $q_{R}$ (Unruh modes) beyond single-mode approximation and implies that the ordering structure employed can give rise to correct physical results. It is found that entanglement has sudden death, therefore quantum coherence and geometric quantum discord are more robust than entanglement in an accelerating system.

\section*{Acknowledgement}
This work is supported by the National Natural Science Foundation of China (Grant No. 61502179),
the Natural Science Foundation of Guangdong Province of China (Grant No. 2014A030310265),
and the Scientific Research Fund for Young Teachers of Wuyi University  (Grant No. 2015zk01).
H.Z. Situ is sponsored by the State Scholarship Fund of the China Scholarship Council.

\end{document}